\documentclass[twocolumn,showpacs,showkeys]{revtex4}
\usepackage{graphicx}
\usepackage{bm}
\usepackage{color}
\usepackage{amsmath}
\usepackage{natbib}

\begin{document}

\title{Kinetic analysis of spin current contribution to spectrum of electromagnetic waves in spin-1/2 plasma, Part II: Dispersion dependencies}
\author{Pavel A. Andreev}
\email{andreevpa@physics.msu.ru}
\affiliation{Faculty of physics, Lomonosov Moscow State University, Moscow, Russian Federation.}
\date{\today}

\begin{abstract}
The dielectric permeability tensor for spin polarized plasmas derived in terms of the spin-1/2 quantum kinetic model in six-dimensional phase space in Part I of this work is applied for study of spectra of high-frequency transverse and transverse-longitudinal waves propagating perpendicular to the external magnetic field.
Cyclotron waves are studied at consideration of waves with electric field directed parallel to the external magnetic field.
It is found that the separate spin evolution modifies the spectrum of cyclotron waves.
These modifications increase with the increase of the spin
polarization and the number of the cyclotron resonance. Spin dynamics
with no account of the anomalous magnetic moment gives a considerable modification of spectra either.
The account of anomalous magnetic moment leads to a fine structure of each cyclotron resonance. So, each cyclotron resonance splits on three waves. Details of this spectrum and its changes with the change of spin polarization are studied for the first and second cyclotron waves.
A cyclotron resonance existing at $\omega\approx0.001\mid\Omega_{e}\mid$ due to the anomalous magnetic moment is also described,
where $\mid\Omega_{e}\mid$ is the cyclotron frequency.
The ordinary waves does not have any considerable modification.
The electrostatic and electromagnetic Berstein modes are studied during the analysis of waves propagating perpendicular to the external magnetic field with the electric field perturbation directed perpendicular to the external field. A modification of the oscillatory structure caused by the equilibrium spin polarization is found in both regimes.
Similar modification is found for the extraordinary wave spectrum.
\end{abstract}


\maketitle




\section{\label{sec:level1} Introduction}

The Fermi spin current existing in the magnetic moment (spin) evolution equation is introduced in Ref. \cite{Andreev 1510 Spin Current}. It is an example of the thermal part of spin current existing for degenerate electrons. Similarity to the pressure, the thermal part of the spin current is an independent hydrodynamic function. It requires an equation of state if we need to make a truncation and obtain a closed set of equation.

In Ref. \cite{Andreev 1510 Spin Current}, an equation of state for the Fermi spin current is found via the derivation of a non-linear Pauli equation from the separate spin evolution Euler equations for the spin-up and spin-down electrons. Found non-linear Pauli equation allows to derive corresponding magnetic moment evolution equation containing explicit form of the Fermi spin current.

Another derivation of the equilibrium Fermi spin current is presented in Ref. \cite{Andreev PoP 16 sep kin}, where the Fermi spin current is derived as a moment of the spin distribution function for degenerate electrons with an arbitrary spin polarization.

Single-fluid spin-1/2 quantum hydrodynamic model of electrons with the Fermi spin current is applied in Ref. \cite{Andreev 1510 Spin Current} for study of spectrum of electromagnetic waves, spin waves and their linear interaction at the wave propagation parallel and perpendicular to the external magnetic field in electron-ion and electron-positron plasmas.
Separate spin evolution quantum hydrodynamics of electrons (two-fluid model of electrons) with the Fermi spin current and the spin-orbit interaction is derived in Ref. \cite{Andreev 1603}, where the extraordinary waves propagating perpendicular to the external magnetic field are studied. The extraordinary spin-electron acoustic wave is found and studied there.

More advanced method of the study of the thermal part of the spin current or the Fermi spin current effect for degenerate gas is a quantum kinetic model containing this effect. Corresponding model is developed in Part I of this paper \cite{Andreev Kinetics DPT pI}. Corresponding dielectric permeability tensor for spin polarized plasmas is derived for the oblique propagating perturbations.

The Fermi spin current is obtained from the separate spin evolution kinetic \cite{Andreev PoP 16 sep kin} can be more relevant for comparison with the kinetic model results. Consider divergence of the Fermi spin current obtained from the kinetic model and find the following vector
$$\mbox{\boldmath $\Im$}_{K}=(6\pi^{2})^{\frac{1}{3}}\frac{\pi\hbar\mu_{e}}{8m}\times$$
\begin{equation}\label{SC_KA spin current many part Vector from Kin}  \times\{n_{\uparrow}^{\frac{1}{3}}\partial_{x} n_{\uparrow}-n_{\downarrow}^{\frac{1}{3}}\partial_{x} n_{\uparrow}, n_{\uparrow}^{\frac{1}{3}}\partial_{y} n_{\uparrow}-n_{\downarrow}^{\frac{1}{3}}\partial_{y} n_{\uparrow}, 0\},\end{equation}
where $n_{\uparrow}$ ($n_{\downarrow}$) is the concentration of electrons in spin-up (spin-down) states, $m$ is the mass of particles, $\mu_{e}$ is the magnetic moment of electron including the anomalous magnetic moment, $\mu_{e}=gq_{e}\hbar/2mc$, $g=1.00116$, $q_{e}$ is the charge of electron $q_{e}=-\mid e\mid$, $c$ is the speed of light, and $\Im_{K}^{\alpha}=\partial_{\beta}J^{\alpha\beta}_{K}$.

The application of the non-linear Pauli equation with the spinor pressure gives another form of the Fermi spin current \cite{Andreev 1510 Spin Current}:
\begin{equation}\label{SC KA spin current many part Vector} \mbox{\boldmath $\Im$}_{P}=\frac{(3\pi^{2})^{\frac{2}{3}}\hbar }{m}(n_{\uparrow}^{\frac{2}{3}}-n_{\downarrow}^{\frac{2}{3}}) [\textbf{M}, \textbf{e}_{z}],\end{equation}
where $\textbf{M}$ is the magnetization of electron gas, $\textbf{e}_{z}$ is the unit vector in the $z$-direction.

The spin-1/2 quantum hydrodynamics containing the Fermi spin current (\ref{SC_KA spin current many part Vector from Kin}) or (\ref{SC KA spin current many part Vector}) (see for instance \cite{Andreev 1510 Spin Current}) is a generalization of the spin-1/2 quantum hydrodynamics suggested earlier \cite{MaksimovTMP 2001, MaksimovTMP 2001 b, Marklund PRL07, Brodin NJP07, Shukla PhUsp 2010, Brodin PRL 10 SPF, Mahajan PRL 11, Shukla RMP 11, Koide PRC 13, Uzdensky RPP 14}.

Equations (\ref{SC_KA spin current many part Vector from Kin}) and (\ref{SC KA spin current many part Vector}) show that the Fermi spin current exists due to the difference of the populations of electrons with different spin projections. However, $\mbox{\boldmath $\Im$}_{K}$ and $\mbox{\boldmath $\Im$}_{P}$ have some difference. $\mbox{\boldmath $\Im$}_{P}$ does not contain any derivatives. Hence, the Fourier image of its linear form does not includes the wave vector. Consequently, $\mbox{\boldmath $\Im$}_{P}$ gives contribution for all directions of wave propagation. On the other hand, $\mbox{\boldmath $\Im$}_{K}$ contains derivatives on space variables. Moreover, it contains derivatives on space directions which are perpendicular to the external magnetic field. Consequently, at the analysis of linear waves, $\mbox{\boldmath $\Im$}_{K}$ does not give any contribution in spectrum of waves propagating parallel to the external magnetic field. Vector $\mbox{\boldmath $\Im$}_{K}$ is derived from the kinetic equations as a moment of the equilibrium spin-distribution function for the spin-polarized degenerate electron gas.


The dielectric permeability tensor is obtained in Part I for the general form of isotropic distribution function. An explicit form of the dielectric permeability tensor is obtained for the spin-1/2 partially polarized three- dimensional electron gas, where the following equilibrium distribution functions are used $f_{0}(p)=[\vartheta(p_{F\uparrow}-p)+\vartheta(p_{F\downarrow}-p)]/(2\pi\hbar)^{3}$ and
$S_{0z}(p)=[\vartheta(p_{F\uparrow}-p)-\vartheta(p_{F\downarrow}-p)]/(2\pi\hbar)^{3}$, $S_{0x}=S_{0y}=0$, where $p$ is the momentum, $p_{Fs}=(6\pi^{2}n_{s})^{\frac{1}{3}}\hbar$ is the Fermi momentum for spin-s electrons, with $s=\uparrow$ or $\downarrow$. It corresponds to the isotropic equilibrium. Spin evolution can be anisotropic $S_{0x}(p,\varphi)$, $S_{0y}(p,\varphi)$ even for isotropic $f_{0}(p)$ and $S_{0z}(p)$. A description of this regime is given in Appendix A of the Part I \cite{Andreev Kinetics DPT pI}.

An anisotropic spin distribution function $\textbf{S}(\textbf{p})$ is used to derive the Fermi spin current (\ref{SC_KA spin current many part Vector from Kin}) via the following definition $J^{\alpha\beta}_{K}=\mu_{e}\int S_{0}^{\alpha}(\textbf{p})v^{\beta}dp$.

Spinless part of the dielectric permeability tensor contains $f_{0}(p)$. The spin dependent part of the dielectric permeability tensor $\varepsilon^{\alpha\beta}(\omega,\textbf{k})$ has two parts. One of them contains $S_{0z}(p)$ and another part is proportional to $f_{0}(p)$. The Fermi spin current effects are related to the difference in occupations of spin-up and spin-down states. Hence, the Fermi spin current effects appear from the terms containing the equilibrium spin distribution function $S_{0z}(p)$.
At the kinetic analysis the Fermi spin current effects exist even for the isotropic distribution functions for all directions of wave propagations, as it is demonstrated in this paper.

Spectrum of waves can be found from the following dispersion equation
\begin{equation}\label{SC_KA disp eq general} det\biggl[k^{2}\delta^{\alpha\beta}-k^{\alpha}k^{\beta} -\frac{\omega^{2}}{c^{2}}\varepsilon^{\alpha\beta}(\omega,\textbf{k})\biggr]=0, \end{equation}
where the dielectric permeability tensor $\varepsilon^{\alpha\beta}(\omega,\textbf{k})$ is obtained in \cite{Andreev Kinetics DPT pI}, $\textbf{k}$ is the wave vector, $k$ is the module of the wave vector, $\omega$ is the frequency of wave, and $c$ is the speed of light. In this paper, the dispersion equation (\ref{SC_KA disp eq general}) is applied for study of the wave spectrum with the Fermi spin current effects.

This paper is organized as follows. In Sec. II is devoted to the waves propagating parallel to the external magnetic field. In Sec. III presents analysis of waves propagating perpendicular to the external magnetic field. In Sec. IV a brief summary of obtained results is presented.

\section{Wave propagation PARALLEL to the external magnetic field}

Considering the wave propagation parallel to the external magnetic field $\textbf{k}=\{0,0,k_{z}\}$, the following dielectric permeability tensor can be derived from equation (52) of the Part I \cite{Andreev Kinetics DPT pI}:
$$\varepsilon^{\alpha\beta}=\delta^{\alpha\beta}-\sum_{s=\uparrow, \downarrow}\int \sin\theta d\theta \biggl[\widetilde{\Pi}^{\alpha\beta}_{Cl}(\theta,s) $$
$$+\frac{m^{2}}{\pi\hbar^{3}}\frac{\mu_{e}^{2}c^{2}}{2\omega^{2}}\sum_{r=+,-} \frac{v_{Fs}^{2}k_{z}\cos\theta\kappa^{\alpha\beta}_{r}}{\omega-k_{z}v_{Fs}\cos\theta+r\Omega_{\mu}}$$
\begin{equation}\label{SC_KA dielectric permeability tensor GF} -\frac{m^{3}}{\pi\hbar^{3}}\frac{\mu_{e}^{2}c^{2}}{\hbar\omega^{2}}\sum_{r=+,-}\int_{0}^{v_{Fs}} \frac{r\kappa^{\alpha\beta}_{r} (-1)^{i_{s}} v^{2}dv}{\omega-k_{z}v \cos\theta+r\Omega_{\mu}}\biggr],\end{equation}
where $\delta^{\alpha\beta}$ is the Kronecker symbol, $\mu_{e}$ is the magnetic moment of particles, $v$ is the module of velocity, $\theta$ is an angle of the spherical coordinates in the velocity space defined as $\cos\theta=v_{z}/v$, $\Omega_{e}=q_{e}B_{0}/mc$, $\Omega_{\mu}=2\mu_{e}B_{0}/\hbar$, $\kappa^{\alpha\beta}_{+}=K^{\alpha\beta}_{\parallel}$, $\kappa^{\alpha\beta}_{-}=(K^{\alpha\beta}_{\parallel})^{*}$, $i_{\uparrow}=0$, $i_{\downarrow}=1$,
\begin{equation}\label{SC_KA } \hat{K}_{\parallel}=k_{z}^{2}\left(\begin{array}{ccc}
1 & -\imath & 0 \\
\imath & 1 & 0 \\
0 & 0 & 0 \\
\end{array}\right),\end{equation}
and
\begin{widetext}
\begin{equation}\label{SC_KA Pi Cl} \widehat{\widetilde{\Pi}}_{Cl}(\theta,s)=\frac{3\omega_{Ls}^{2}}{2\omega}\left(\begin{array}{ccc}
\frac{1}{4}(\frac{\sin^{2}\theta}{\omega-k_{z}v_{Fs}\cos\theta-\Omega_{e}}+\frac{\sin^{2}\theta}{\omega-k_{z}v_{Fs}\cos\theta+\Omega_{e}}) & \imath \frac{1}{4}(\frac{\sin^{2}\theta}{\omega-k_{z}v_{Fs}\cos\theta-\Omega_{e}}-\frac{\sin^{2}\theta}{\omega-k_{z}v_{Fs}\cos\theta+\Omega_{e}}) & 0 \\
-\imath \frac{1}{4}(\frac{\sin^{2}\theta}{\omega-k_{z}v_{Fs}\cos\theta-\Omega_{e}}-\frac{\sin^{2}\theta}{\omega-k_{z}v_{Fs}\cos\theta+\Omega_{e}}) & \frac{1}{4}(\frac{\sin^{2}\theta}{\omega-k_{z}v_{Fs}\cos\theta-\Omega_{e}}+\frac{\sin^{2}\theta}{\omega-k_{z}v_{Fs}\cos\theta+\Omega_{e}}) & 0 \\
0 & 0 & \frac{\cos^{2}\theta}{\omega-k_{z}v_{Fs}\cos\theta}
\end{array}\right),\end{equation}
where well-known properties of the Bessel functions are used, and $v_{Fs}=p_{Fs}/m=(6\pi^{2}n_{0s})^{1/3}\hbar/m$, $\omega_{Ls}^{2}=4\pi e^{2}n_{0s}/m$. $\widehat{\widetilde{\Pi}}_{Cl}(\theta,s)$ is similar to the traditional result for degenerate electrons presented in many textbooks (see for instance \cite{Rukhadze book 84}), but it also includes the spin separation effect. The third and fourth terms in equation (\ref{SC_KA dielectric permeability tensor GF}) are caused by the spin evolution via the dynamics of the spin-distribution function.

Further integration in the dielectric permeability tensor gives the following result:
$$\varepsilon^{\alpha\beta}=\delta^{\alpha\beta}-\sum_{s=\uparrow, \downarrow} \Biggl[\widetilde{\Pi}^{\alpha\beta}_{Cl}(s) +\frac{m^{2}v_{Fs}}{\pi\hbar^{3}}\frac{\mu_{e}^{2}c^{2}}{2\omega^{2}}
\sum_{r=+,-}\kappa_{r}^{\alpha\beta} \Biggl(-2+\frac{\omega +r\Omega_{\mu}}{k_{z}v_{Fs}} \ln\biggl(\frac{\omega+k_{z}v_{Fs}+r\Omega_{\mu}}{\omega-k_{z}v_{Fs}+r\Omega_{\mu}}\biggr)\Biggr)$$
\begin{equation}\label{SC_KA } +\frac{m^{3}}{\pi\hbar^{3}}\frac{\mu_{e}^{2}c^{2}}{\hbar\omega^{2}} \sum_{r=+,-}r\kappa^{\alpha\beta}_{r}(-1)^{i_{s}}\Biggl(-\frac{v_{Fs}}{k_{z}^{2}}(\omega+r\Omega_{\mu}) +\frac{(\omega+r\Omega_{\mu})^{2}-(k_{z}v_{Fs})^{2}}{2k_{z}^{3}} \ln\biggl(\frac{\omega+k_{z}v_{Fs}+r\Omega_{\mu}}{\omega-k_{z}v_{Fs}+r\Omega_{\mu}}\biggr)\Biggr)\Biggr]\end{equation}
with 
\begin{equation}\label{SC_KA Pi Cl} \widehat{\widetilde{\Pi}}_{Cl}(s)=\frac{3\omega_{Ls}^{2}}{2\omega}\left(\begin{array}{ccc}
\frac{1}{4}[G(\omega-\Omega_{e})+G(\omega+\Omega_{e})] & \imath \frac{1}{4}[G(\omega-\Omega_{e})-G(\omega+\Omega_{e})] & 0 \\
-\imath \frac{1}{4}[G(\omega-\Omega_{e})-G(\omega+\Omega_{e})] & \frac{1}{4}[G(\omega-\Omega_{e})+G(\omega+\Omega_{e})] & 0 \\
0 & 0 & \frac{\omega}{(k_{z}v_{Fs})^{2}}(-2+\frac{\omega}{k_{z}v_{Fs}}\ln(\frac{\omega+k_{z}v_{Fs}}{\omega-k_{z}v_{Fs}}))
\end{array}\right),\end{equation}\end{widetext}
where
$$G(\omega\mp\Omega_{e})=
\frac{1}{k_{z}v_{Fs}}\biggl[\frac{2(\omega\pm\Omega_{e})}{k_{z}v_{Fs}}$$
\begin{equation}\label{SC_KA }
+\biggl(1-\frac{(\omega\pm\Omega_{e})^{2}}{(k_{z}v_{Fs})^{2}}\biggr)\ln\biggl(\frac{\omega+k_{z}v_{Fs}\pm\Omega_{e}}{\omega-k_{z}v_{Fs}\pm\Omega_{e}}\biggr)\biggr]. \end{equation}

The dispersion equation appears in the following form at the application of the found structure of the dielectric permeability tensor:
\begin{equation}\label{SC_KA }det\left(
\begin{array}{ccc}
    k_{z}^{2}-\frac{\omega^{2}}{c^{2}}\varepsilon_{xx} & -\frac{\omega^{2}}{c^{2}}\varepsilon_{xy} & 0 \\
    -\frac{\omega^{2}}{c^{2}}\varepsilon_{yz} & k_{z}^{2}-\frac{\omega^{2}}{c^{2}}\varepsilon_{yy} & 0 \\
    0 & 0 & -\frac{\omega^{2}}{c^{2}}\varepsilon_{zz} \\
\end{array}\right)=0.\end{equation}

It splits on three equations, one equation is for the longitudinal waves
\begin{equation}\label{SC_KA varepsilon zz=0} \varepsilon_{zz}=0,\end{equation}
and two equations are for the transverse waves
\begin{equation}\label{SC_KA tr waves general form} \frac{k_{z}^{2}c^{2}}{\omega^{2}}-\epsilon=\pm\Xi,\end{equation}
where it is used that $\varepsilon_{xx}=\varepsilon_{yy}\equiv\epsilon$, and $\varepsilon_{yx}^{*}=\varepsilon_{xy}=\imath\Xi$.

Explicit form of dispersion equation (\ref{SC_KA varepsilon zz=0}) appears as follows:
\begin{equation}\label{SC_KA tr waves par pr expl form} 1+\sum_{s=\uparrow, \downarrow}\frac{3\omega_{Ls}^{2}}{k_{z}^{2}v_{Fs}^{2}} \Biggl(1-\frac{1}{2}\frac{\omega}{k_{z}v_{Fs}}\ln\biggl(\frac{\omega+k_{z}v_{Fs}}{\omega-k_{z}v_{Fs}}\biggr)\Biggr)=0.\end{equation}

Consideration of the spin-polarized equilibrium distribution functions as the two Fermi step functions gives same dispersion equation for the longitudinal waves as in the separate spin evolution kinetics \cite{Andreev PoP 16 sep kin}.

Equation (\ref{SC_KA tr waves par pr expl form}) reveals existence of the spin-electron acoustic waves and their Landau damping \cite{Andreev PoP 16 sep kin}. The bulk spin-electron acoustic wave propagating parallel to the external magnetic field is found theoretically in terms of the separate spin evolution quantum hydrodynamics \cite{Andreev PRE 15 SEAW}. Existence of two spin-electron acoustic waves in the regime of oblique wave propagation is demonstrated in \cite{Andreev AoP 15 SEAW}. Reacher spectrum of the spin-electron acoustic waves is found in electron-positron-ion plasmas \cite{Andreev PRE 16}, \cite{Andreev_Iqbal PoP 16}. A method of account of the Coulomb exchange interaction in the separate spin evolution quantum hydrodynamics is developed in \cite{Andreev PoP 16 exchange} for more detailed analysis of linear and non-linear properties of
spin-electron acoustic waves. The surface spin-electron acoustic waves are studied either \cite{Andreev APL 16}. A possibility of the linear interaction between the surface plasmon and the surface spin-electron acoustic wave is found. An applications of methods of the condensed matter physics \cite{Agarwal PRL 11}, \cite{Agarwal PRB 14} allows to find similar effects in graphine \cite{Agarwal PRB 15}, \cite{Sachdeva PRB 15}.

Dispersion equations for the transverse waves appear from (\ref{SC_KA tr waves general form}) as follows
$$\frac{k_{z}^{2}c^{2}}{\omega^{2}}=1 -\sum_{s=\uparrow, \downarrow}
\Biggl[\frac{3}{4}\frac{\omega_{Ls}^{2}}{\omega} \frac{1}{k_{z}v_{Fs}}\biggl[\frac{2(\omega\pm\Omega_{e})}{k_{z}v_{Fs}}$$
\begin{equation}\label{SC_KA tr waves explicit form} +\biggl(1-\frac{(\omega\pm\Omega_{e})^{2}}{(k_{z}v_{Fs})^{2}}\biggr) \ln\biggl(\frac{\omega+k_{z}v_{Fs}\pm\Omega_{e}}{\omega-k_{z}v_{Fs}\pm\Omega_{e}}\biggr)\biggr] +C_{s}+D_{s}\Biggr],\end{equation}
for $\delta E_{x}=\pm\imath\delta E_{y}$ correspondingly, where
$$C_{s}(\pm)=k_{z}^{2}\frac{m^{2}v_{Fs}}{\pi\hbar^{3}}\frac{\mu_{e}^{2}c^{2}}{\omega^{2}}\Biggl(-2+\frac{\omega\pm\Omega_{\mu}}{k_{z}v_{Fs}}\times $$ $$\times\ln\biggl(\frac{\omega+k_{z}v_{Fs}\pm\Omega_{\mu}}{\omega-k_{z}v_{Fs}\pm\Omega_{\mu}}\biggr)\Biggr),$$
and
$$D_{s}(\pm)=(-1)^{i_{s}}\frac{m^{3}}{\pi\hbar^{3}}\frac{\mu_{e}^{2}c^{2}}{\hbar\omega^{2}}\Biggl(-2v_{Fs}(\omega\pm\Omega_{\mu})$$
\begin{equation}\label{SC_KA tr waves explicit form D s} +\frac{1}{k_{z}}\biggl((\omega\pm\Omega_{\mu})^{2}-(k_{z}v_{Fs})^{2}\biggr) \ln\biggl(\frac{\omega+k_{z}v_{Fs}\pm\Omega_{\mu}}{\omega-k_{z}v_{Fs}\pm\Omega_{\mu}}\biggr)\Biggr),\end{equation}
where $C_{s}(\pm)$ is caused by the spin evolution and appears via $f_{0}$, and $D_{s}(\pm)$ is also caused by the spin evolution, but $D_{s}(\pm)$ appears via $S_{0z}$. Below, an approximate analytical analysis shows that $D_{s}(\pm)$ contains effects of spin polarization $\sim n_{0\uparrow}-n_{0\downarrow}$ which are similar to the spin evolution in hydrodynamics. $D_{s}(\pm)$ also contains effects caused by $((1-\eta)^{\frac{5}{3}}-(1+\eta)^{\frac{5}{3}})$ looking like difference of Fermi pressures which are similar to the Fermi spin current.
$C_{s}(\pm)$ and $D_{s}(\pm)$ exist even for neutral particles. Kinetic model of spin evolution in system of neutral particles is presented in Ref. \cite{Andreev Phys A 15}.

Dispersion equation (\ref{SC_KA tr waves explicit form}) describes the circularly polarized transverse waves with two polarizations. Consider these equations in the long-wavelength limit $\omega\gg kv_{Fs}$:
$$\frac{k_{z}^{2}c^{2}}{\omega^{2}}=1 -\sum_{s=\uparrow, \downarrow}\Biggl[\frac{\omega_{Ls}^{2}}{\omega k_{z}v_{Fs}}\biggl[\frac{k_{z}v_{Fs}}{\omega\mp\mid\Omega_{e}\mid}
+\frac{1}{5}\biggl(\frac{k_{z}v_{Fs}}{\omega\mp\mid\Omega_{e}\mid}\biggr)^{3}\biggr]$$
$$+\frac{\omega_{Ls}^{2}}{\omega^{2}}\frac{\hbar^{2}k_{z}^{2}}{m^{2}v_{Fs}^{2}}\biggl(\frac{k_{z}v_{Fs}}{\omega\mp\mid\Omega_{\mu}\mid}\biggr)^{2}$$
\begin{equation}\label{SC_KA tr waves explicit form expanded} -(-1)^{i_{s}} \frac{\omega_{Ls}^{2}}{\omega^{2}}\frac{\hbar k_{z}}{2mv_{Fs}} \biggl[\frac{k_{z}v_{Fs}}{\omega\mp\mid\Omega_{\mu}\mid}
+\frac{1}{5}\biggl(\frac{k_{z}v_{Fs}}{\omega\mp\mid\Omega_{\mu}\mid}\biggr)^{3}\biggr]\Biggr],\end{equation}
for $\delta E_{x}=\pm\imath\delta E_{y}$. Equation (\ref{SC_KA tr waves explicit form expanded}) contains coefficients proportional to $\hbar k_{z}/m$. It resembles a similarity to the well-known quantum Bohm potential. However, it comes from the spin evolution \cite{Maksimov VestnMSU 2000}, \cite{Andreev VestnMSU 2007}, \cite{Misra JPP 10}. That shows that spectrum is modified at large $k$ due to the spin dynamics. Separated evolution of spin-up and spin-down electrons does not modify the degree of equation (\ref{SC_KA tr waves explicit form expanded}). Hence, it does not change number of transverse waves propagating parallel to the external magnetic field. Equation (\ref{SC_KA tr waves explicit form expanded}) shows that the spin evolution mostly contributes in dispersion equation at the large wave vectors. A shift of the imaginary part of $\omega(k)$ for right-hand circularly polarized waves propagating parallel to the external magnetic field at large $k$ due to the spin dynamics is demonstrated in \cite{Iqbal PoP 14}. The last term in equation (\ref{SC_KA tr waves explicit form expanded}) has contribution similar to the Fermi spin current.

Equation (\ref{SC_KA tr waves explicit form expanded}) contains the equation which is well-known from spin-1/2 hydrodynamics:
\begin{equation}\label{SC_KA tr waves explicit form expanded  reduced}
\frac{k_{z}^{2}c^{2}}{\omega^{2}}=1
-\frac{\omega_{Le}^{2}}{\omega (\omega\mp\mid\Omega_{e}\mid)}
+\frac{\omega_{Le}^{2}}{\omega^{2}}\frac{\hbar k_{z}^{2}}{2m n_{0e}} \frac{n_{0\uparrow}-n_{0\downarrow}}{\omega\mp\mid\Omega_{\mu}\mid}
,\end{equation}
see for instance \cite{Misra JPP 10}.
In equation (\ref{SC_KA tr waves explicit form expanded  reduced}) $\omega_{Le}^{2}=\omega_{L\uparrow}^{2}+\omega_{L\downarrow}^{2}$ is the Langmuir for all electrons and $n_{0e}=n_{0\uparrow}+n_{0\downarrow}$ is the concentration of all electrons.
An equation similar to equation (\ref{SC_KA tr waves explicit form expanded  reduced}), but for waves propagating perpendicular to the external magnetic field, containing resonances near double cyclotron frequency $2\mid\Omega_{e}\mid$ and near $\mid\Omega_{\mu}-\Omega_{e}\mid$ are found in \cite{Zamanian PoP 10} via an extended hydrodynamics including spin velocity moment evolution equation.


Deriving equation (\ref{SC_KA tr waves explicit form expanded}) from equation (\ref{SC_KA tr waves explicit form}), the expansion of ln function in the Taylor series is considered up to the fifth order on $kv_{Fs}/(\omega\mp\mid\Omega_{e}\mid)$. Traditionally this expansion is restricted by the third order (see for instance \cite{Rukhadze book 84}). The spin effects in the last regime exist via the expansion of $D_{s}(\pm)$ (see equation (\ref{SC_KA tr waves explicit form expanded  reduced})).
In equation (\ref{SC_KA tr waves explicit form expanded}) it exists as $\sum_{s}(-1)^{i_{s}} \frac{\omega_{Ls}^{2}}{\omega^{2}}\frac{\hbar k_{z}}{2mv_{Fs}} \frac{k_{z}v_{Fs}}{\omega\mp\mid\Omega_{\mu}\mid}\sim n_{0\uparrow}-n_{0\downarrow}$. The next order of the expansion is considered to include the pressure-like term which is proportional to $(-1)^{i_{s}}v_{Fs}^{2}$. This term describes the spin evolution with the account of the Fermi spin current.

Equation (\ref{SC_KA tr waves explicit form expanded}) shows that the effects of the Fermi spin current are small in the limit $kv_{Fs}/(\omega\mp\mid\Omega_{e}\mid)\ll1$, since they appear in the highest order on the small parameter.

\section{Wave propagation PERPENDICULAR to the external magnetic field}

This section is devoted to the wave propagation perpendicular to the external magnetic field, $\textbf{k}=\{k_{x},0,0\}$.
In this regime the dielectric permeability tensor given by equation (52) of the Part I \cite{Andreev Kinetics DPT pI} modifies to
$$\varepsilon^{\alpha\beta}=\delta^{\alpha\beta}-\sum_{s=\uparrow, \downarrow} \biggl\{\sum_{n=-\infty}^{\infty}\biggl[\frac{\int \sin\theta d\theta\widetilde{\Lambda}^{\alpha\beta}(n,s)}{\omega-n\Omega_{e}} $$
$$ +\frac{m^{2}v_{Fs}}{\pi\hbar^{3}} \frac{\mu_{e}^{2}c^{2}}{2\omega^{2}} K^{\alpha\beta}_{\perp} \sum_{r=+,-}
\frac{ n\Omega_{e}\int \sin\theta d\theta J_{n}^{2} }{ \omega-n\Omega_{e}+r\Omega_{\mu} } $$
$$-\frac{m^{3}}{\pi\hbar^{3}} \frac{\mu_{e}^{2}c^{2}}{\hbar\omega^{2}} K^{\alpha\beta}_{\perp} (-1)^{i_{s}}\sum_{r=+,-}
\frac{r}{\omega-n\Omega_{e}+r\Omega_{\mu}}\times$$
$$\times \int_{0}^{v_{Fs}} v^{2}dv \int \sin\theta d\theta J_{n}^{2}\biggl(\frac{k_{x}v\sin\theta}{\Omega_{e}}\biggr)\biggr]$$
\begin{equation}\label{SC_KA dielectric permeability tensor GF par} -2\mu_{e}^{2}\frac{m^{2}v_{Fs}}{\pi\hbar^{3}}\frac{k_{x}^{2}c^{2}}{\omega^{2}}  \delta^{\alpha y}\delta^{\beta y}\biggr\},\end{equation}
where $\sum_{n=-\infty}^{+\infty}J_{n}^{2}=1$, $\sum_{n=-\infty}^{+\infty}J_{n}J_{n}'=0$ are used in the last group of terms of equation (52) of the Part I \cite{Andreev Kinetics DPT pI}, and
$$\widetilde{\Lambda}^{\alpha\beta}(n,s)=\frac{3\omega_{Ls}^{2}}{2\omega v_{Fs}^{2}}\Pi^{\alpha\beta}_{Cl}(n,s) +m^{2}v_{Fs}\frac{\mu_{e}^{2}}{\pi\hbar^{3}}\frac{k_{x}^{2}c^{2}}{\omega}J_{n}^{2}\delta^{\alpha y}\delta^{\beta y} $$
\begin{equation}\label{SC_KA } +\frac{q_{e}\mu_{e}}{\pi\hbar^{3}}\frac{c}{\omega}m^{2}v_{Fs}(-1)^{i_{s}}\Pi^{\alpha\beta}_{S}(n,s),\end{equation}
with                          
$$\widehat{\Pi}_{Cl}(n,s)=v_{Fs}^{2}\times$$
\begin{equation}\label{SC_KA Pi Cl} \times\left(\begin{array}{ccc}
\frac{\Omega_{e}^{2}}{k_{x}^{2}v_{Fs}^{2}}n^{2} J_{n}^{2} & \imath\sin\theta\frac{\Omega_{e}}{k_{x}v_{Fs}} nJ_{n}J_{n}' & 0 \\
-\imath\sin\theta\frac{\Omega_{e}}{k_{x}v_{Fs}} nJ_{n}J_{n}' & \sin^{2}\theta(J_{n}')^{2} & 0 \\
0 & 0 & \cos^{2}\theta J_{n}^{2}
\end{array}\right)\end{equation}                
which has structure similar to well-known from textbooks \cite{Rukhadze book 84}, but it separately describes electrons with spin-up and spin-down,
and the tensor
\begin{equation}\label{SC_KA } \Pi^{\alpha\beta}_{S}(n,s)=\left(
\begin{array}{ccc}
0 & \imath\Omega_{e}nJ_{n}^{2} & 0 \\
-\imath\Omega_{e}nJ_{n}^{2} & 2k_{x}v_{Fs}\sin\theta J_{n}J_{n}' & 0 \\
0 & 0 & 0 \\
\end{array}\right)\end{equation}
describes the spin evolution leading to the same resonances as the classic evolution $\omega=k_{z}v_{Fs}\cos\theta+n\Omega_{e}$.
Here, the Bessel functions $J_{n}$ are functions of $k_{x}v_{Fs}\sin\theta/\mid\Omega_{e}\mid$, if a Bessel function has different argument this argument is shown explicitly as it is in the third line in equation (\ref{SC_KA dielectric permeability tensor GF par}), and $J_{n}'$ means derivative of $J_{n}$ on its argument $k_{x}v_{Fs}\sin\theta/\mid\Omega_{e}\mid$.
Two spin dependent terms in equation (\ref{SC_KA dielectric permeability tensor GF par}) contain the following matrix
\begin{equation}\label{SC_KA } \hat{K}_{\perp}=k_{x}^{2}\left(
\begin{array}{ccc}
0 & 0 & 0 \\
0 & 0 & 0 \\
0 & 0 & 1\\
\end{array}\right).\end{equation}

Similarly to the spinless case, the dispersion equation in the regime of wave propagation perpendicular to the external magnetic field splits on two independent equations:
\begin{equation}\label{SC_KA disp eq perp ordinary waves} \frac{k_{x}^{2}c^{2}}{\omega^{2}}=\varepsilon_{zz}\end{equation}
for the transverse waves with the linear polarization $\delta \textbf{E}=\{0,0,\delta E_{z}\}$ (ordinary waves), and
\begin{equation}\label{SC_KA disp eq perp longitudinally-transverse waves} \frac{k_{x}^{2}c^{2}}{\omega^{2}}=\frac{\varepsilon_{xx}\varepsilon_{yy}+\varepsilon_{xy}^{2}}{\varepsilon_{xx}}\end{equation}
for the longitudinally-transverse (extraordinary) waves.

Equations (\ref{SC_KA disp eq perp ordinary waves}) and (\ref{SC_KA disp eq perp longitudinally-transverse waves}) includes the dispersion dependence of the cyclotron and Bernstein modes. Analysis of the longitudinal Bernstein modes in relativistic and non-relativistic regimes of classic plasmas with comparable cyclotron and Langmuir frequencies can be found in \cite{Keston PRE 04}.

\begin{figure}
\includegraphics[width=8cm,angle=0]{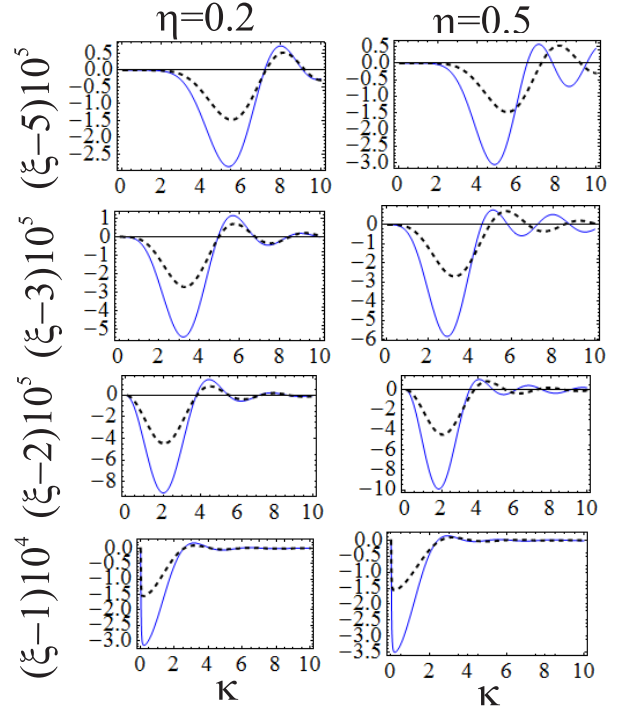}
\caption{\label{susdSOI 05} (Color online) The figure shows the influence of the separate spin evolution on spectrum of the first, second, third and fifth "ordinary" cyclotron waves for $\eta=0.2$ (the left-hand side column) and $\eta=0.5$ (the right-hand side column). The separate spin evolution enters via the traditional terms which are not caused by the spin dynamics.} \end{figure}
\begin{figure}
\includegraphics[width=8cm,angle=0]{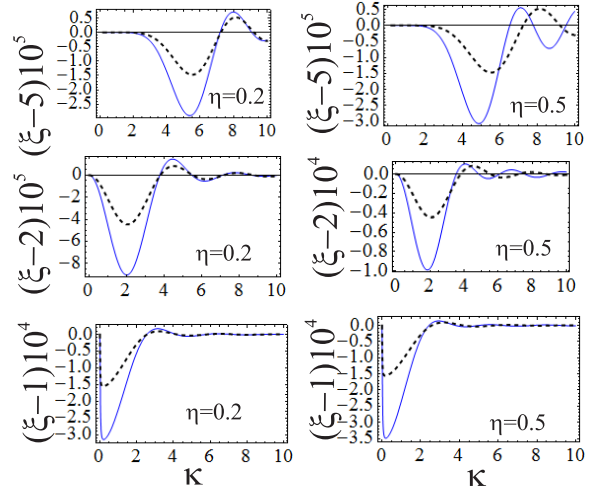}
\caption{\label{susdSOI 05} (Color online) The figure shows the influence of the spin dynamics on the spectrum of under assumption $\Omega_{\mu}=\Omega_{e}$ made in equation (\ref{SC_KA disp eq perp ordinary waves explicit 1}) for spectrum of the first, second and fifth "ordinary" cyclotron waves for $\eta=0.2$ (the left-hand side column) and $\eta=0.5$ (the right-hand side column). In this regime the number of roots is not changed, but spectrum is affected by the spin dynamics.} \end{figure}
\begin{figure}
\includegraphics[width=8cm,angle=0]{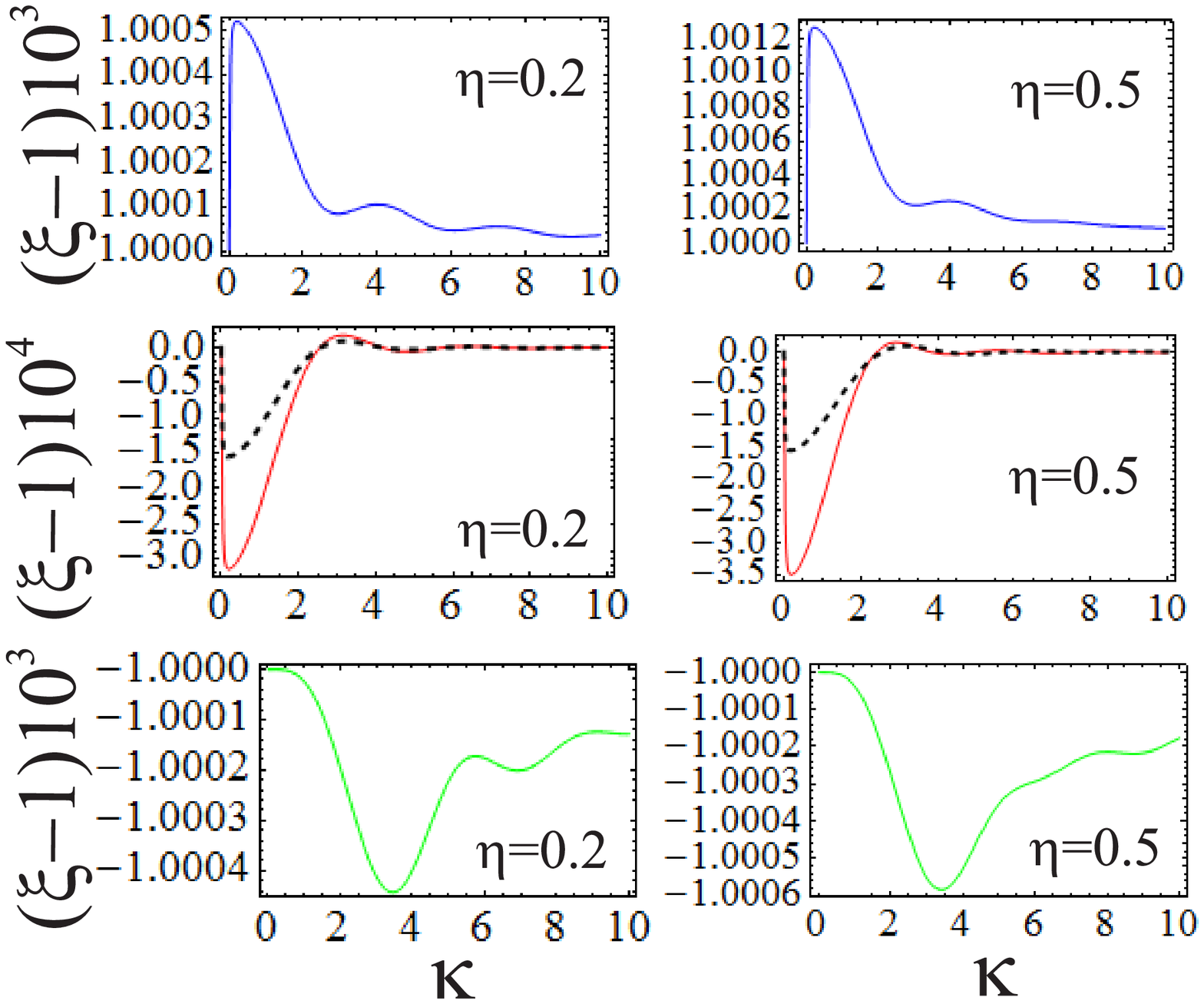}
\caption{\label{susdSOI 05} (Color online) The figure shows details of the fine structure of the first cyclotron mode. It is shown for two values of the spin polarization.} \end{figure}
\begin{figure}
\includegraphics[width=8cm,angle=0]{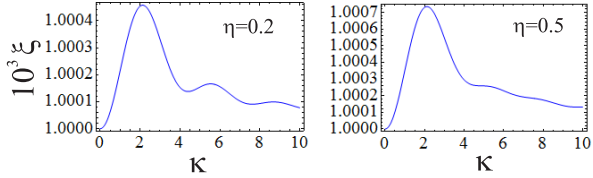}
\caption{\label{susdSOI 05} (Color online) The figure shows the dispersion curve for the lowest cyclotron wave (the zeroth cyclotron wave). This wave exist due to the spin evolution and has frequency near $0.001 \mid\Omega_{e}\mid$. It is shown for two values of the spin polarization.} \end{figure}
\begin{figure}
\includegraphics[width=8cm,angle=0]{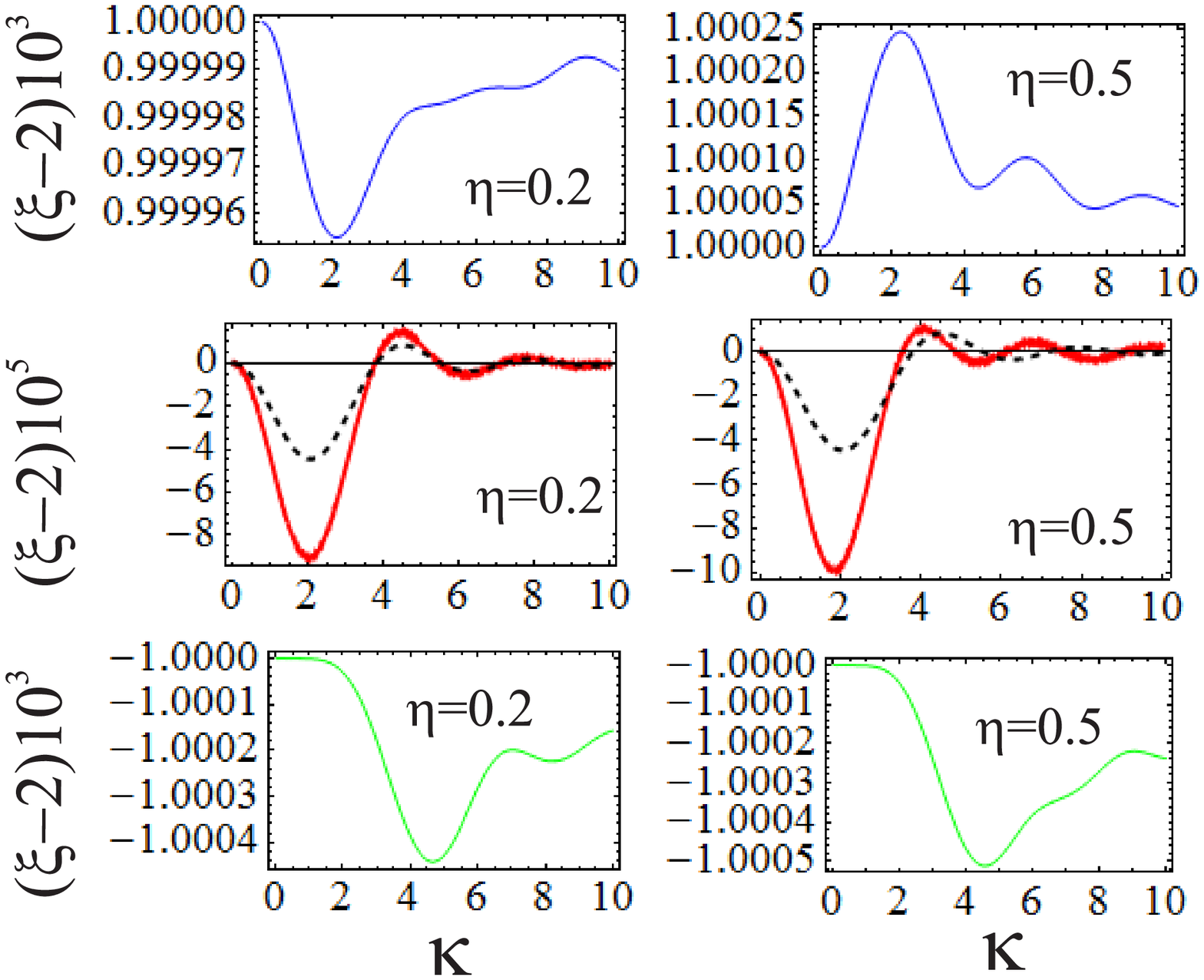}
\caption{\label{susdSOI 05} (Color online) The figure shows details of the fine structure of the second cyclotron mode. It is shown for two values of the spin polarization.} \end{figure}
\begin{figure}
\includegraphics[width=8cm,angle=0]{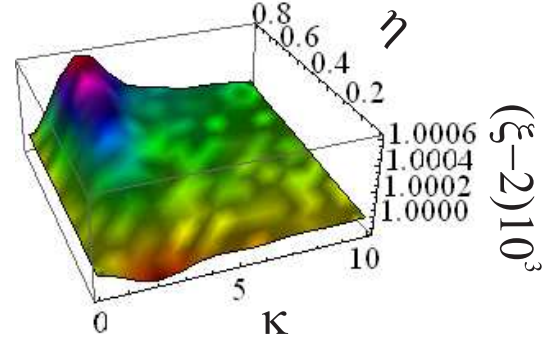}
\caption{\label{susdSOI 05} (Color online) The figure shows the modification of the dispersion curve of the upper line in the triplet of the second cyclotron mode with the change of the spin polarization.} \end{figure}

\subsection{Linearly polarized waves: Ordinary waves, cyclotron waves and spin waves}

Consideration of the ordinary wave spectrum requires a detailed analysis of equation (\ref{SC_KA disp eq perp ordinary waves}).
To this end, an explicit form of equation (\ref{SC_KA disp eq perp ordinary waves}) is presented as follows
$$\frac{k_{x}^{2}c^{2}}{\omega^{2}}=1-\sum_{s=\uparrow, \downarrow} \sum_{n=-\infty}^{\infty} \biggl[\frac{3\omega_{Ls}^{2}}{2\omega}\frac{\int \sin\theta \cos^{2}\theta J_{n}^{2} d\theta}{\omega-n\Omega_{e}}$$
$$ +\frac{m^{2}v_{Fs}}{\pi\hbar^{3}}\frac{\mu_{e}^{2}c^{2}}{2\omega^{2}} k_{x}^{2}\sum_{r=+,-}\frac{n\Omega_{e}\int \sin\theta d\theta J_{n}^{2}}{\omega-n\Omega_{e}+r\Omega_{\mu}} -\frac{(-1)^{i_{s}}m^{3}}{\pi\hbar^{3}}\frac{\mu_{e}^{2}c^{2}}{\hbar\omega^{2}}\times$$
\begin{equation}\label{SC_KA disp eq perp ordinary waves explicit 1} \times k_{x}^{2}\sum_{r=+,-}r
\frac{\int_{0}^{v_{Fs}} v^{2}dv \int \sin\theta d\theta J_{n}^{2}(\frac{k_{x}v\sin\theta}{\mid\Omega_{e}\mid})}{\omega-n\Omega_{e}+r\Omega_{\mu}}\biggr]. \end{equation}
Equation (\ref{SC_KA disp eq perp ordinary waves explicit 1}) describes the ordinary waves including the ordinary cyclotron waves. The last two terms in equation (\ref{SC_KA disp eq perp ordinary waves explicit 1}) are caused by the dynamics of $\delta S_{x}$ and $\delta S_{y}$.
The last term in equation (\ref{SC_KA disp eq perp ordinary waves explicit 1}) contains the contribution of spin-up and spin-down electrons with different signs. Hence, the Fermi spin current effects enter the ordinary cyclotron wave spectrum via the last term.

All nontrivial terms on the right-hand side of equation (\ref{SC_KA disp eq perp ordinary waves explicit 1}) contain effects of the separate spin evolution. Except the separate spin evolution, the second term has same form as in traditional spinless case. The second term (which is proportional to $\omega_{Ls}^{2}$) has resonances on harmonics of the electron cyclotron frequency. The third and fourth terms are caused by the spin dynamics. The resonances in the third and fourth terms are shifted by the magnetic moment cyclotron frequency $\pm\Omega_{\mu}$. The spin dynamics leads to new resonances and formation of the fine structure of the cyclotron waves. It is similar to the Bernstein modes structure found in Ref. \cite{Hussain PP 14 spin bernst}.
However, Ref. \cite{Hussain PP 14 spin bernst} deals with the electrostatic Bernstein modes with the electric field directed along $0x$ axis $\delta \textbf{E}=\{\delta E_{x}, 0, 0\}$. This regime is considered in the next subsection. Here, the ordinary transverse cyclotron waves are under consideration. Presented here results have similarity with Ref. \cite{Hussain PP 14 spin bernst}, but this is a different phenomenon.
The effect of splitting of the electrostatic Bernstein modes is demonstrated in \cite{Hussain PP 14 spin bernst} for the second Bernstein mode which includes $\omega=2\mid\Omega_{e}\mid$, $\omega=3\mid\Omega_{e}\mid-\mid\Omega_{\mu}\mid$, $\omega=\mid\Omega_{e}\mid+\mid\Omega_{\mu}\mid$. The splitting gives three closely located dispersion curves (a triplet) since $\Omega_{\mu}\neq\Omega_{e}$ and $\mid\Omega_{\mu}-\Omega_{e}\mid\ll\mid\Omega_{e}\mid$. The Fermi-Dirac distribution modified by a factor describing the spin polarization is applied in Ref. \cite{Hussain PP 14 spin bernst} as an equilibrium distribution function.

Considering different waves it would be unnecessary comparing the dispersion equations. However, there are some similarities which can be mentioned. The first term on the right-hand side of equation 26 of Ref. \cite{Hussain PP 14 spin bernst} is a classic term. The second and third terms on the right-hand side of equation 26 of Ref. \cite{Hussain PP 14 spin bernst} appear due to the spin dynamics. The second and third terms are actually two terms of sum taken at $n=1$ and $n=3$. So, the right-hand side of equation 26 of Ref. \cite{Hussain PP 14 spin bernst} consists of two groups of terms: classical and spin terms.
The second term on the right-hand side of equation (\ref{SC_KA disp eq perp ordinary waves explicit 1}) is a classic term. The third term gives in-phase contribution of spin-up and spin down states and resemblance a distant similarity to second group of terms in equation 26 of Ref. \cite{Hussain PP 14 spin bernst}. The fourth term in equation (\ref{SC_KA disp eq perp ordinary waves explicit 1}) presents antiphase contribution of spin-up and spin down states and has no analogs in equation 26 of Ref. \cite{Hussain PP 14 spin bernst}.

Solve equation (\ref{SC_KA disp eq perp ordinary waves explicit 1}) in several regimes. First, dropping the spin dynamics (the third and fourth terms on the right-hand side), study the spectrum change due to the separate spin evolution. Second, include all terms, but drop the anomalous part of the magnetic moment (it gives $\Omega_{e}=\Omega_{\mu}$). Third, consider all effects presented in equation (\ref{SC_KA disp eq perp ordinary waves explicit 1}). All regimes require numerical solution of equation (\ref{SC_KA disp eq perp ordinary waves explicit 1}).

Results of solution of equation (\ref{SC_KA disp eq perp ordinary waves explicit 1}) in the first regime are presented in Fig. 1.
Find the following considering the first regime. Fig. 1 shows that the separate spin evolution effect becomes larger for higher cyclotron waves. Fig. 1 shows that an increase of the spin polarization $\eta$ leads to a larger modification of spectrum. The numerical analysis in this subsection is made for $n_{0}=10^{21}$ cm$^{-3}$ and $B_{0}=10^{7}$ G. This is regime of dense plasma with $\omega_{Le}\gg\Omega_{e}$.

First of all this modification appears at relatively large wave vectors $\textbf{k}$. It appears via faster oscillation of the dispersion curve. Moreover, the divergency of the dispersion curve from $n\Omega_{e}$ decreases. As is is well-known, the cyclotron waves and Bernstein modes appears in kinetic theory due to the detailed description of the distribution function evolution in the momentum space. Hydrodynamic description of these effects requires account of the higher moments of the distribution function, such as pressure evolution, which allows to get the Bernstein mode near $2\mid\Omega_{e}\mid$ \cite{Zamanian PoP 10}. The spin polarization modifies the Fermi step to two different steps for the spin-up and spin-down electrons. Expectedly, it modifies the Bernstein mode spectrum.

The second regime reveals the following modification of spectrum of the cyclotron waves.
The spin dynamics changes form of spectrum of cyclotron waves (see Fig. 2). Particularly, the divergence of a curve from $n\mid\Omega_{e}\mid$ (for corresponding $n$) increases several times.

The account of anomalous magnetic moment leads to the fine structure of the cyclotron waves. Each curve splits on three closely located curves Fig. 3. Resonances near $n\mid\Omega_{e}\mid$ appears due to the following denominators: $\omega-n\mid\Omega_{e}\mid$, $\omega-(n+1)\mid\Omega_{e}\mid+\mid\Omega_{\mu}\mid$, $\omega-(n-1)\mid\Omega_{e}\mid-\mid\Omega_{\mu}\mid$.

Considering $n=0$, find one resonance at a positive frequency $\omega-(\mid\Omega_{\mu}\mid-\mid\Omega_{e}\mid)=\omega-0.001\mid\Omega_{e}\mid$. Solution of equation (\ref{SC_KA disp eq perp ordinary waves explicit 1}) confirms that there is a branch near $0.001\mid\Omega_{e}\mid$ (see Fig. 4). This effect is demonstrated in \cite{Brodin PRL 08 g Kin} in terms of kinetic model. It is also found in terms of extended quantum hydrodynamics containing equations for the pressure moment and the spin-velocity moment \cite{Zamanian PoP 10}.
A spin modified Maxwellian distribution is employed in Ref. \cite{Brodin PRL 08 g Kin}. So, they are consider a non-degenerate plasma. As a result the dispersion curve $\omega(k)$ monotonically increases, reach maximum value, and monotonically decay. While Figs. 3 and 5 shows that there are oscillations of $\omega(k)$ after reaching the maximum value. It is in accordance with the oscillations of the dispersion curves of cyclotron and Bernstein modes in degenerate plasmas.
Here, this wave found as a part of spectrum of ordinary waves, which is in accordance with Ref. \cite{Brodin PRL 08 g Kin}.

The spin dynamics contributions in the dielectric permeability tensor appear to be proportional to one of the following dimensionless constants: $e^{2}n_{0e}^{1/3}/mc^2$ and $mcen_{0e}^{1/3}/\hbar B_{0}\sim e^{2}n_{0e}^{1/3}/\hbar\Omega_{e}$. They present the ratio of the average Coulomb interaction energy to the rest energy of electron, and the ratio of the average Coulomb interaction energy to the uanta of cyclotron oscillation, correspondingly.

There is a change of the triplet structure of the first cyclotron wave due to the spin polarization change (see Fig. 3). Same conclusion is correct for the zeroth cyclotron wave (see Fig. 4). However, there is a modification of dispersion curve of the upper line of the triplet of the second cyclotron wave. The modification of the upper curve due to the spin polarization change is demonstrated in
Fig. 6. Change of the light and dark areas shows the oscillatory structure of the dispersion curve and its change. the monotonic increase of the maximal deviation with the increase of the spin polarization is demonstrated either.

The upper line in the triplet of the first cyclotron wave has a deviation towards larger frequencies. The middle and lower lines of the triplet of the first cyclotron wave are deviated towards smaller frequencies. The increase of the spin polarization makes the oscillatory structure of curves is less noticeable, but maximal deviation increases for all waves in the triplet. Similar picture is found for the zeroth cyclotron wave (see Fig. 4).

The upper and middle lines in the triplet of the second cyclotron wave have reacher oscillatory structure at the larger spin polarization. It is in opposite to the behavior of the triplet of the first cyclotron wave. The lower line in the triplet of the second cyclotron wave shows more similarity to the triplet of the first cyclotron wave and has smaller oscillatory structure at the large spin polarization. The maximal deviation of the middle and lower lines in the triplet of the second cyclotron wave increases with the increase of the spin polarization, similarity to the triplet of the first cyclotron wave.

Calculation of the imaginary part of frequencies for considered in this subsection waves shows that they have no collisionless damping.

\begin{figure}
\includegraphics[width=8cm,angle=0]{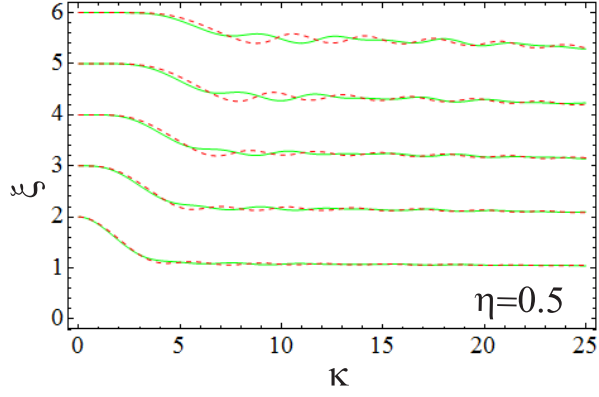}
\caption{\label{susdSOI 05} (Color online) The figure shows the electrostatic Bernstein modes in two regimes. The dashed red lines show the classic spectrum existing in degenerate electron gas with no spin polarization. The continuous green lines show the spectrum with account of the separate spin evolution.
It is plotted for $n_{0}=10^{21}$ cm$^{-3}$ and $B_{0}=10^{7}$ G.} \end{figure}
\begin{figure}
\includegraphics[width=8cm,angle=0]{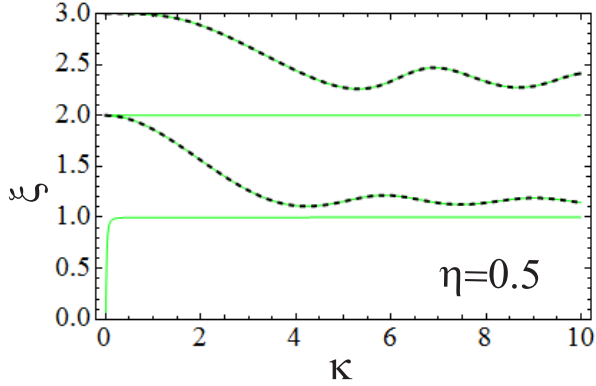}
\caption{\label{susdSOI 05} (Color online) The figure shows the dispersion curves for Bernstein modes with account of the transverse electric field and the separate spin evolution (green continuous lines) and dispersion curves for Bernstein modes in the electrostatic regime with the account of separate spin evolution (black dashed curves).} \end{figure}
\begin{figure}
\includegraphics[width=8cm,angle=0]{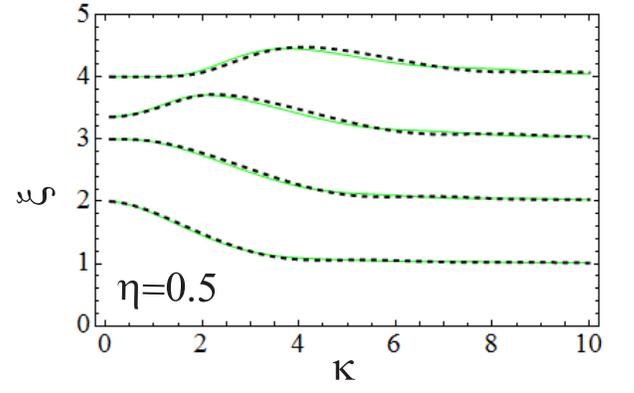}
\caption{\label{susdSOI 05} (Color online) The figure shows a small modification of the spectrum of the extraordinary wave at the account of the separate spin evolution. The green continuous lines show spectra of the Bernstein modes and extraordinary wave without of the separate spin evolution. The black dashed lines present the spectra with the account of the separate spin evolution. The figure is plotted for the following parameters: $n_{0}=10^{19}$ cm$^{-3}$ and $B_{0}=10^{7}$ G.} \end{figure}

\subsection{Elliptically polarized modes: Bernstein modes, extraordinary wave}

This subsection is devoted to the longitudinally-transverse waves. Neglecting the transverse part they reduces to the electrostatic modes: Langmuir (hybrid) wave, Bernstein modes.

The dielectric permeability tensor elements contained in equation (\ref{SC_KA disp eq perp longitudinally-transverse waves}) have the following form:
\begin{equation}\label{SC_KA e xx perp 1} \varepsilon_{xx}=1-\sum_{s=\uparrow, \downarrow} \sum_{n=1}^{\infty} \frac{n^{2}\Omega_{e}^{2}}{k_{x}^{2}v_{Fs}^{2}} \frac{3\omega_{Ls}^{2}}{\omega^{2}-n^{2}\Omega_{e}^{2}}\int \sin\theta J_{n}^{2} d\theta, \end{equation}
$$\varepsilon_{yy}=1-\sum_{s=\uparrow, \downarrow} \Biggl\{\sum_{n=1}^{\infty} \frac{1}{\omega^{2}-n^{2}\Omega_{e}^{2}}\Biggl[3\omega_{Ls}^{2} \int \sin^{3}\theta (J_{n}')^{2}d\theta$$
$$+2 \frac{m^{2}\mu_{e}^{2}}{\pi\hbar^{3}}k_{x}^{2}c^{2}v_{Fs} \int \sin\theta J_{n}^{2}d\theta$$
$$ +4(-1)^{i_{s}}\frac{q_{e}\mu_{e}}{\pi\hbar^{3}}m^{2}v_{Fs}^{2}k_{x}c
\int \sin^{2}\theta J_{n}J_{n}'d\theta \Biggr]
+2\frac{m^{2}v_{Fs}}{\pi\hbar^{3}}\mu_{e}^{2}\frac{k_{x}^{2}c^{2}}{\omega^{2}}$$
$$+\frac{3\omega_{Ls}^{2}}{2\omega^{2}}\int \sin^{3}\theta (J_{0}')^{2}d\theta
+\frac{\mu_{e}^{2}}{\pi\hbar^{3}}\frac{k_{x}^{2}c^{2}}{\omega^{2}} m^{2}v_{Fs} \int \sin\theta J_{0}^{2}d\theta$$
\begin{equation}\label{SC_KA e yy perp 1} +2(-1)^{i_{s}}\frac{q_{e}\mu_{e}}{\pi\hbar^{3}}m^{2}v_{Fs}^{2}\frac{k_{x}c}{\omega^{2}}
\int \sin^{2}\theta J_{0}J_{0}'d\theta \Biggr\},\end{equation}
and
$$\varepsilon_{xy}= -\sum_{s=\uparrow, \downarrow} \sum_{n=1}^{\infty} \frac{2n^{2}\Omega_{e}^{2}}{\omega^{2}-n^{2}\Omega_{e}^{2}} \biggl[\frac{3\omega_{Ls}^{2}}{2\omega}\frac{1}{k_{x}v_{Fs}} \times $$
\begin{equation}\label{SC_KA e xy perp 1} \times\int d\theta \sin^{2}\theta J_{n}J_{n}' +(-1)^{i_{s}}\frac{m^{2}v_{Fs}}{\omega}\frac{q_{e}\mu_{e}c}{\pi\hbar^{3}} \int d\theta \sin\theta J_{n}^{2}\biggr].\end{equation}

$\varepsilon_{xx}$ is the longitudinal dielectric permeability.
Equation (\ref{SC_KA e xx perp 1}) shows that the spin dynamics gives no contribution in $\varepsilon_{xx}$. The separate spin evolution related to the equilibrium spin polarization modifies $\varepsilon_{xx}$ in compare with the classic model.
$\varepsilon_{yy}$ and $\varepsilon_{xy}$ are modified by the spin dynamics. The spin related terms  are proportional to $\mu_{e}$. Hence, they can be easily identified.
All nontrivial terms in $\varepsilon_{xx}$, $\varepsilon_{xy}$ and $\varepsilon_{yy}$ are proportional to $(\omega^{2}-n^{2}\Omega_{e}^{2})^{-1}$, with n=0, 1, 2, ... since equations (\ref{SC_KA e xx perp 1})-(\ref{SC_KA e xy perp 1}) are caused by dynamics of $\delta f$ and $\delta S_{z}$.
$\Omega_{\mu}$ does not enter the denominators in equations (\ref{SC_KA e xx perp 1})-(\ref{SC_KA e xy perp 1}). So, there is no fine structure of the Bernstein modes appearing as solutions of equation (\ref{SC_KA disp eq perp longitudinally-transverse waves}).
The triplet fine structure of the cyclotron waves is found in the previous subsection. The triplet fine structure of the Bernstein modes is found in Ref. \cite{Hussain PP 14 spin bernst} in terms of another form of spin-1/2 kinetic model, but it does not appear in the described model.

In the electrostatic limit the dispersion equation reads $\varepsilon_{xx}=0$. Several solutions of this equation are shown in Fig. 7. The separate spin evolution modifies the oscillatory structure of the Bernstein modes. The modification becomes more prominent with the increase of the mode number. Next, include the transverse electric field. To this end, the dispersion equation (\ref{SC_KA disp eq perp longitudinally-transverse waves}) is solved, where $\varepsilon_{xx}$, $\varepsilon_{xy}$ and $\varepsilon_{yy}$ are considered in accordance with equations (\ref{SC_KA e xx perp 1})-(\ref{SC_KA e yy perp 1}) dropping terms proportional to the magnetic moment $\mu_{e}$. Corresponding results are presented in Fig. 8. It shows that the spectrum of Bernstein modes is not changed, but additional lines appear. Finally, the terms caused by the spin dynamics are to include. In this regime, the full elements of the dielectric permeability tensor presented by equations (\ref{SC_KA e xx perp 1})-(\ref{SC_KA e yy perp 1}) are employed. Comparison of coefficients in classic and spin terms shows a relative increase of the spin terms at the large wave vector.

Modification of the extraordinary wave spectrum due to the separate spin evolution is found and presented in Fig. 9. It is rather small, but noticeable contribution. The spin dynamics in considered regime gives even smaller contribution which is not presented in figures.

This subsection shows that the major contribution in the Bernstein modes and extraordinary wave spectra in the considered parameter regime gives the separate spin evolution of electrons.

\section{Conclusion}

Two groups of results have been found. One is related to the transverse linearly polarized (ordinary) waves propagating perpendicular to the external magnetic field with the electric field perturbation parallel to the external magnetic field.
The second group of results is related to the longitudinally-transverse elliptically polarized (extraordinary) waves propagating perpendicular to the external magnetic field with the electric field perturbation perpendicular to the external magnetic field.

The following results are found in the first regime. The spectrum modification of the cyclotron waves and the triplet fine structure for each cyclotron wave existing due to combination of spin effects have been demonstrated.
Properties of the lowest spin-cyclotron wave with the frequency near $0.001 \mid\Omega_{e}\mid$ have been studied for the partially spin polarized degenerate electron gas.
The anomalous magnetic moment leads to the fine structure formation and appearance of the lowest cyclotron wave. However, several effects contribute into the form of dispersion curves. They are the classic terms modified by the separate spin evolution and terms caused by the spin dynamics (they also contain the separate spin evolution). The spin terms are specific in two ways. They give a noticeable contribution even if the anomalous magnetic moment is neglected, but the anomalous magnetic moment causes additional effects like the fine structure. Relative contributions of these effects are traced analytically and numerically.

The Bernstein modes are considered in the second regime. It has been found that the separate spin evolution modifies the oscillatory structure of the Bernstein modes. The modification increases with the increase of the mode number. This effects remains at the account of the transverse field. Similar modification is found for the extraordinary wave.

First of all the obtained results are important for the magnetically ordered conductors and semiconductors. They can be applied to the degenerate astrophysical plasmas with the large spin polarization of charge carriers either.

\begin{acknowledgements}
The author thanks Professor L. S. Kuz'menkov for fruitful discussions. The work was supported by the Russian
Foundation for Basic Research (grant no. 16-32-00886) and the Dynasty foundation.
\end{acknowledgements}


\end{document}